\documentclass[twocolumn,showpacs,preprintnumbers]{revtex4}
\usepackage{amssymb}
\usepackage{graphics}
\usepackage{graphicx}

\begin{document}

\title{Spectral function and fidelity susceptibility in quantum critical
phenomena}
\date{\today }

\begin{abstract}
In this paper, we derive a simple equality that relates the spectral
function $I(k,\omega)$ and the fidelity susceptibility $\chi_F$, i.e. $%
\chi_F=\lim_{\eta\rightarrow 0}\frac{\pi}{\eta} I(0, i\eta)$ with $\eta$
being the half-width of the resonance peak in the spectral function. Since
the spectral function can be measured in experiments by the neutron
scattering or the angle-resolved photoemission spectroscopy(ARPES)
technique, our equality makes the fidelity susceptibility directly
measurable in experiments. Physically, our equality reveals also that the
resonance peak in the spectral function actually denotes a quantum
criticality-like point at which the solid state seemly undergoes a
significant change.
\end{abstract}

\author{Shi-Jian Gu}
\email{sjgu@phy.cuhk.edu.hk}
\affiliation{Department of Physics and ITP, The Chinese University of Hong Kong, Hong
Kong, China}
\author{Wing Chi Yu}
\affiliation{Department of Physics and ITP, The Chinese University of Hong Kong, Hong
Kong, China}
\pacs{05.30.Rt, 75.40.Gb, 64.60.-i, 03.67.-a}
\maketitle

%75.40.Gb Dynamic properties (dynamic susceptibility, spin waves, spin
%diffusion, dynamic scaling, etc.)

%05.30.Rt 	Quantum phase transitions (see also 64.70.Tg Quantum phase transitions in specific phase transitions; and 73.43.Nq Quantum phase transitions in Quantum Hall effects)

%03.67.-a    quantum information
%64.60.-i    General studies of phase transitions (see also 63.70.+h
%Statistical mechanics of lattice vibrations and displacive phase transitions;
%for critical phenomena in solid surfaces and interfaces, and in
%magnetism, see 68.35.Rh, and 75.40.-s, respectively)
%75.10.-b General theory and models of magnetic ordering (see also
%05.50.+q Lattice theory and statistics)

At zero temperature, the ground-state properties of a quantum many-body
system can change quantitatively as the system's parameter varies across a
critical point. Because of the absence of thermal fluctuation, the
quantitative change is solely driven by quantum fluctuation and hence is
called a quantum phase transition \cite{Sachdev}. Examples of quantum phase
transitions include Mott-insulator transitions and fractional quantum Hall
liquids. In the perspective of the quantum information science \cite%
{Nielsen1}, the ground-state wavefunctions on both sides of the critical
point $\lambda _{c}$ have distinct structures and if we compare two ground
states separated by a small fixed distance $\delta \lambda $ in the
parameter space, i.e. the fidelity $|\langle \psi _{0}(\lambda )|\psi
_{0}(\lambda +\delta \lambda )\rangle |$, is expected to show a minimum at
the critical point $\lambda _{c}$ \cite{ZanardiPRE,HQZhouJPA2008}. The
quantum phase transition in the perspective of the fidelity have been
verified in many strongly correlated systems \cite%
{PZanardi0606130,MCozzini07,MCozzini072,Buonsante1}. On the other hand,
since the structure of the ground-state wavefunction undergoes a significant
change as the system is driven adiabatically across the transition point, we
can also imagine that the leading term of the fidelity, i.e. the fidelity
susceptibility which denotes the leading response of the ground state to the
driving parameter, should be a maximum or even divergent at the transition
point \cite{WLYou2007,PZanardi2007G}. Besides, the fidelity between two
ground states separated by a long distance in the parameter space also
manifests distinct information about quantum phase transitions \cite%
{HQZhouPRL2008,BogdanPRL2011}. Due to the remarkable properties of the
fidelity around the critical point \cite%
{BogdanPRL2011,SChenPRA2008,VenutiPRL2007,SJGuPRB2008}, the fidelity has
become an efficient way to detect the quantum transition point in quantum
many-body systems \cite%
{SJGUReview,AHammaPRB2008,JHZhouPRB2009,SYangPRA2008,DFAbasto2008,MFYangFidelity,HMKwok2008PRE,JieRenPRA,GreschnerPRB2013,Oliveira2014,WLYou2PRB,XuebingPRA2014}%
. Especially, the fidelity has proven to be able to detect unconventional
phase transitions such as the topological phase transition too \cite%
{AHammaPRB2008,JHZhouPRB2009,SYangPRA2008,DFAbasto2008}.

Despite of the great success of the fidelity approach to quantum phase
transitions in theory, little progress has been achieved in experiments. Up
to now, the only experimental detection of the quantum phase transition in
terms of fidelity is based on a spin dimer system via the technique of the
nuclear-magnetic-resonance quantum simulator \cite{ZhangPRLexp}. For a large
quantum many-body system, say having a size $L>10$, to measure the overlap
of its two ground states separated by a short distance in the parameter
space seems hard to be realized. The interesting scaling and universality
behaviors of the fidelity susceptibility in quantum phase transitions still
cannot been verified in experiments. Therefore, it is highly expected to
find a way to measure the fidelity and its susceptibility directly or
indirectly in experiments.

In this paper, we finally derive a neat equality that connects two
seemingly unrelated quantities, i.e. the spectral function and fidelity
susceptibility. Since the spectral function can be measured in experiments
by such as the neutron scattering or ARPES technique \cite{ARPESbook}, such
an equality actually makes the fidelity susceptibility directly measurable
in experiments. On the other hand, as the most typical model in quantum
phase transitions, the transverse-field Ising model and its quantum
criticality now can be studied in experiment via the neutron scattering \cite%
{RColdeal}. A possible experimental scheme to measure the fidelity
susceptibility of the transverse-field Ising model is proposed.

To begin with, we consider the propagation properties of a single electron
in a solid-state system. Without the loss of generality, we assume that the
system can be described by a Hubbard-like model whose Hamiltonian reads
\begin{equation}
H=-t\sum_{\langle ij\rangle \sigma }c_{i,\sigma }^{\dagger }c_{j,\sigma
}+U\sum_{j}n_{j,\uparrow }n_{j,\downarrow }+H_{V},  \label{eq:Hamiltonian}
\end{equation}%
where $\left\langle {}\right\rangle $ denotes the summation over all the
nearest neighboring pairs, $c_{j,\sigma }^{\dagger }$($c_{j,\sigma }$) is
the creation(annihilation) operator for electrons with spin $\sigma
=\uparrow ,\downarrow $ at site $j$, $t$ is the hoping integral, $%
n_{j,\sigma }=c_{j,\sigma }^{\dagger }c_{j,\sigma }$, $U$ is the strength of
on-site Coulomb interaction, and $H_{V}$ denotes other types of
interactions. In a solid-state system, the total number of electrons is a
good quantum number and is decided by the chemical potential of the system.
Let us assume that the sample system has $N$ electrons. In this subspace,
the eigenstates of the system are decided by the Schr\"{o}dinger equation%
\begin{equation}
H|\psi _{n}^{N}\rangle =E_{n}|\psi _{n}^{N}\rangle .
\end{equation}%
At zero temperature, the propagation of a single electron in the ground
state $|\psi _{0}^{N}\rangle $ can be described by the one-electron Green's
function in the momentum-energy space%
\begin{equation}
G^{\pm }(k,\omega )=\sum_{m}\frac{|\left\langle \psi _{m}^{N\pm 1}\left\vert
c_{k}^{\pm }\right\vert \psi _{0}^{N}\right\rangle |^{2}}{\omega
+E_{0}^{N}-E_{m}^{N\pm 1}+i\eta },  \label{eq:GreenFk}
\end{equation}%
where $|\psi _{m}^{N\pm 1}\rangle $ is the eigenstate of the Hamiltonian in
the subspace of $N\pm 1$ electrons and $c_{k}^{+}=c_{k\sigma }^{\dagger
}(c_{k}^{-}=c_{k\sigma })$,%
\begin{equation}
c_{k\sigma }^{\dagger }=\frac{1}{\sqrt{V}}\sum_{j}e^{-ijk}c_{j,\sigma
}^{\dagger }
\end{equation}%
with $V$ being the volume of the system. The one-electron spectral function%
\begin{equation}
I^{\pm }(k,\omega )=-\frac{1}{\pi }\mathrm{Im}G^{\pm }(k,\omega )
\end{equation}%
defines the one-electron addition and removal spectra. $I^{\pm }(k,\omega )$
can be probed in the inverse and direct photoemission, respectively \cite%
{ARPESbook}. From Eq. (\ref{eq:GreenFk}), we have
\begin{equation}
I^{\pm }(k,\omega )=\lim_{\eta \rightarrow 0}\sum_{m}\frac{\eta }{\pi }\frac{%
|\left\langle \psi _{m}^{N\pm 1}\left\vert c_{k}^{\pm }\right\vert \psi
_{0}^{N}\right\rangle |^{2}}{\left( \omega +E_{0}^{N}-E_{m}^{N\pm 1}\right)
^{2}+\eta ^{2}}.  \label{eq:specdensiy}
\end{equation}%
We notice the right-hand side of Eq. (\ref{eq:specdensiy}) actually denotes
the leading response of the ground state in the photoemission process.
Precisely, the form of $I^{\pm }(k,\omega )$ is already the same as the
dynamic fidelity susceptibility introduced in Ref. \cite{WLYou2007}. To
observe this, we need to consider the effective Hamiltonian including the
subspace of both $N-1$ and $N+1$ electrons,
\begin{equation}
H(\eta )=\left(
\begin{array}{ccc}
H(N-1)-\omega  & \eta c_{k}^{+} &  \\
\eta c_{k}^{-} & H(N) & \eta c_{k}^{+} \\
& \eta c_{k}^{-} & H(N+1)-\omega
\end{array}%
\right)
\end{equation}%
where\ $\omega $ is due to the photon absorbtion and emission and $\eta $ is
the strength of the perturbation. When $\eta =0$, the initial ground state
of the system $|\psi _{0}^{N}\rangle $ locates in the subspace of $N$
electrons. Then if a small perturbation $\eta (c_{k}^{+}+c_{k}^{-})$ is
turned on, the state becomes, to the first order,
\begin{eqnarray}
|\psi _{0}(\eta )\rangle  &=&\left\vert \psi _{0}^{N}\right\rangle +\eta
\sum_{m}\frac{\langle \psi _{m}^{N+1}|c_{k}^{+}\left\vert \psi
_{0}^{N}\right\rangle |\psi _{m}^{N+1}\rangle }{\omega +E_{0}^{N}-E_{m}^{N+1}%
}  \nonumber \\
&&+\eta \sum_{m}\frac{\langle \psi _{m}^{N-1}|c_{k}^{-}\left\vert \psi
_{0}^{N}\right\rangle |\psi _{m}^{N-1}\rangle }{\omega +E_{0}^{N}-E_{m}^{N-1}%
}.  \label{eq:persfftestate}
\end{eqnarray}%
According to the definition \cite{ZanardiPRE}, the fidelity between $%
\left\vert \psi _{0}^{N}\right\rangle $ and $|\psi _{0}(\eta )\rangle $
becomes%
\begin{equation}
|\langle \psi _{0}^{N}|\psi _{0}(\eta )\rangle |=1-\frac{\eta ^{2}}{2}\chi
_{F}+\cdots
\end{equation}%
where%
\begin{eqnarray}
\chi _{F} &=&\sum_{m}\frac{|\langle \psi _{m}^{N+1}|c_{k}^{+}\left\vert \psi
_{0}^{N}\right\rangle |^{2}}{\left( \omega +E_{0}^{N}-E_{m}^{N+1}\right) ^{2}%
}  \nonumber \\
&&+\sum_{m}\frac{|\langle \psi _{m}^{N-1}|c_{k}^{-}\left\vert \psi
_{0}^{N}\right\rangle |^{2}}{\left( \omega +E_{0}^{N}-E_{m}^{N-1}\right) ^{2}%
}
\end{eqnarray}%
is the so-called fidelity susceptibility. In Ref. \cite{WLYou2007}, we
introduced the concept of dynamic fidelity susceptibility as%
\begin{eqnarray}
\chi _{F}(\eta ) &=&\sum_{m}\frac{|\langle \psi
_{m}^{N+1}|c_{k}^{+}\left\vert \psi _{0}^{N}\right\rangle |^{2}}{\left(
\omega +E_{0}^{N}-E_{m}^{N+1}\right) ^{2}+\eta ^{2}}  \nonumber \\
&&+\sum_{m}\frac{|\langle \psi _{m}^{N-1}|c_{k}^{-}\left\vert \psi
_{0}^{N}\right\rangle |^{2}}{\left( \omega +E_{0}^{N}-E_{m}^{N-1}\right)
^{2}+\eta ^{2}}.
\end{eqnarray}%
Since $I(k,\omega )=I^{+}(k,\omega )+I^{-}(k,\omega )$, compare the above
equation with Eq. (\ref{eq:specdensiy}), we obtain the following equality
\begin{equation}
I(k,\omega )=\lim_{\eta \rightarrow 0}\frac{\eta }{\pi }\chi _{F}(\eta ),
\label{eq:specinfidelitysus}
\end{equation}%
or the inverse
\begin{equation}
\chi _{F}=\lim_{\eta \rightarrow 0}\frac{\pi }{\eta }I(k,\omega +i\eta ),
\label{eq:specinfidelitysus2}
\end{equation}%
which is the key result of this work.

The equality about the fidelity susceptibility and the spectral function is
remarkable. The former is a quantum information theoretic concept used to
study quantum phase transitions. Physically, the divergence of the fidelity
susceptibility manifests a significant change occurred in the structure of
the ground-state wavefunction, hence denotes a phase transition. Lots of
attentions have been paid to the fidelity and fidelity susceptibility
approach to quantum phase transitions in recent years \cite{SJGUReview}.
Nevertheless, the corresponding experimental verification proved to be
extremely difficult \cite{ZhangPRLexp}. Eq. (\ref{eq:specinfidelitysus})
provides us a feasible way to measure the fidelity susceptibility in
experiments via the neutron scattering or ARPES technique. Therefore, the
equality makes the fidelity approach to quantum criticality not merely a
theoretical topic. On the other hand, Eq. (\ref{eq:specinfidelitysus})
reveals that the resonance peak in the spectral function denotes a quantum
criticality-like point at which the solid state of the sample system seemly
undergoes a significant change. Such an interpretation provides us a new
angle to understand the spectral function from the viewpoint of quantum
information science.

In Ref. \cite{WLYou2007}, when we defined the concept of dynamic fidelity
susceptibility $\chi _{F}(\eta )$, the variable $\eta $ was introduced
solely for mathematical purpose to due with the Fourier transformation in
the complex plane. Since then, no work has ever touched the further meaning
of $\eta $. In the equality in Eq. (\ref{eq:specinfidelitysus}), we find it
is $\eta $ that connects the two seemingly unrelated quantities from two
distinct fields. Moreover, $\eta $ appends more physical understanding from
the equality. From the definition of the fidelity susceptibility, $\eta $ is
the strength of the perturbation. While in the definition of the spectral
function, $\eta $ actually denotes the half-width of the resonance peaks.
The uncertainty property of $\eta $ even relates to the lifetime of the
quasi-particle of the resonance peak.

To see the role of the equality in Eq. (\ref{eq:specinfidelitysus}) in
quantum phase transitions, in the following, we take the one-dimensional
transverse-field Ising model as an example to show how to measure the
fidelity susceptibility in experiments. The model's Hamiltonian reads%
\begin{equation}
H=-\sum_{j=1}^{N}\left( \sigma _{j}^{z}\sigma _{j+1}^{z}+h\sigma
_{j}^{x}\right) ,
\end{equation}%
where $\sigma _{j}^{x,y,z}$ is the Pauli Matrix for the 1/2-spin at site $j$%
, $h$ is the transverse field, and $N$ is the number of spins. The periodic
boundary conditions are assumed. The ground state of the Ising model has two
distinct phases, which are the ferromagnetic phase favored by the
ferromagnetic Ising interaction in the Hamiltonian and the paramagnetic
phase due to the transverse field along $+x$ direction. The competition
between them leads to a quantum phase transition occurring at $h_{c}=1$.

The one-dimensional quantum Ising model can be realized by several
materials. An excellent material is the insulating Ising ferromagnet CoNb$%
_{2}$O$_{6}$ whose spin dynamics can be measured by neutron scattering \cite%
{RColdeal}. The model is defined on the zigzag structure formed by Co$^{2+}$
ions whose ferromagnetic coupling is about 1 meV (according to a magnetic
field of 10T $\sim $ 1meV). Then the critical field of the systems is about $%
h=5.5$T which is attainable in the laboratory\cite{RColdeal}.

Theoretically, the model can be diagonalized by the Jordan-Wigner
transformation
\begin{equation}
\sigma _{j}^{z}=1-2c_{j}^{\dagger }c_{j},\text{ \ }\sigma
_{j}^{+}=\prod_{n<j}\sigma _{n}^{z}c_{j},  \label{eq:JW-}
\end{equation}%
the Fourier transformation
\begin{equation}
c_{j}=\frac{1}{\sqrt{N}}\sum_{k}e^{-ikj}c_{k},
\end{equation}%
and the Bogoliubov transformation%
\begin{equation}
c_{k}=u_{k}b_{k}+iv_{k}b_{-k}^{\dagger },
\end{equation}%
where $c_{j}$ is the annihilation operator for spinless fermions at site $j$%
. With the diagonalization condition
\begin{eqnarray}
u_{k}^{2}-v_{k}^{2} &=&\cos 2\theta _{k}=\frac{-(\cos k-h)}{\sqrt{(\cos
k-h)^{2}+\sin ^{2}k}}, \\
2u_{k}v_{k} &=&\sin 2\theta _{k}=\frac{-\sin k}{\sqrt{(\cos k-h)^{2}+\sin
^{2}k}},
\end{eqnarray}%
the Hamiltonian becomes%
\begin{equation}
H=\sum_{k}\epsilon (k)\left( 2b_{k}^{\dagger }b_{k}-1\right) ,
\label{eq:FinalHamIsing}
\end{equation}%
where $b_{k}$ and $b_{k}^{\dagger }$ are fermionic operators, and $\epsilon
(k)=\sqrt{1-2h\cos (k)+h^{2}}$ is the dispersion relation of the
quasi-particles of $b_{k}^{\dagger }$.

\textit{Fidelity approach:} From Eq. (\ref{eq:FinalHamIsing}), we see that
the ground state of the system is the vacuum state of $b_{k}^{\dagger }$,
i.e. $b_{k}|\psi _{0}\rangle =0$. Since $b_{k}(b_{k}^{\dagger })$ depends on
the driving parameter $h$, to compare the two ground states, we should
define them in the space of $c_{k}$ and $c_{k}^{\dagger }$. The ground state
can then be written as
\[
|\psi _{0}\rangle =\prod_{k>0}\left( \cos \theta _{k}|0_{k},0_{-k}\rangle
+i\sin \theta _{k}|1_{k},1_{-k}\rangle \right) ,
\]%
where $|1_{k}\rangle =c_{k}^{\dagger }|0_{k}\rangle $. Under the same basis,
the fidelity between two ground states becomes%
\[
F(h,h^{\prime })=\left\vert \langle \psi _{0}(h)|\psi _{0}(h^{\prime
})\rangle \right\vert =\prod_{k>0}\cos (\theta _{k}-\theta _{k}^{\prime }).
\]%
The fidelity susceptibility, as the leading term in the expansion of $%
F(h,h^{\prime })$, can be calculated explicitly \cite{ZanardiPRE} as%
\begin{equation}
\chi _{F}=\sum_{k>0}\left( \frac{d\theta _{k}}{dh}\right) ^{2},
\label{eq:fsisingana}
\end{equation}%
where
\begin{equation}
\frac{d\theta _{k}}{dh}=\frac{1}{2}\frac{\sin k}{1-2h\cos k+h^{2}}.
\end{equation}

\begin{figure}[tbp]
\includegraphics[width=10cm]{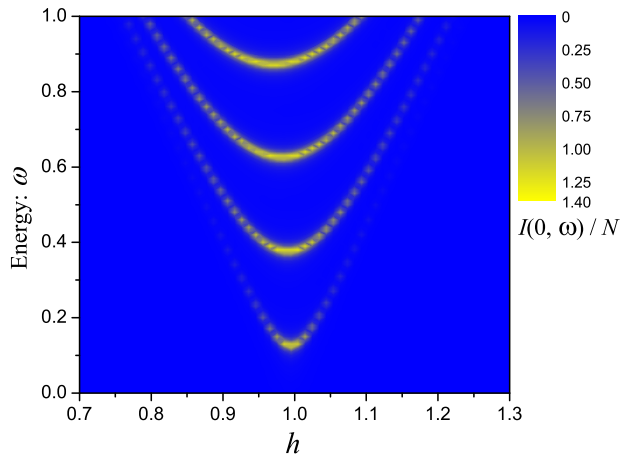}
\caption{(Color online) The spectral function $I(k=0,\protect\omega)$ as a
function of $\protect\omega$ and $h$. Here $N=100$ and $\protect\eta=0.01$.}
\label{figurepec.eps}
\end{figure}

\textit{Spectral function:} In order to measure the fidelity susceptibility
in experiments, we introduce
\begin{equation}
\sigma _{k}^{\pm }=\sum_{j}e^{\mp ijk}\sigma _{j}^{\pm },
\end{equation}%
where $\sigma _{j}^{+}+\sigma _{j}^{-}=\sigma _{j}^{x}$ . Let $A_{k}=\sigma
_{k}^{+}+\sigma _{k}^{-}$, then $A_{k=0}=\sum_{j=1}^{N}\sigma _{j}^{x}$ is
just the driving term of the Hamiltonian. The spectral function of $A_{k}$
can be written as
\begin{equation}
I(k,\omega +i\eta )=\sum_{n}\left\langle \psi _{n}\left\vert
A_{k}\right\vert \psi _{0}\right\rangle ^{2}\delta (E_{0}-E_{n}+\omega
+i\eta ).  \label{eq:specralfuudqs}
\end{equation}%
Let $k=0$,
\begin{equation}
A_{0}=N-2\sum_{k}\big[(u_{k}^{2}-v_{k}^{2})b_{k}^{\dagger
}b_{k}+iu_{k}v_{k}(b_{k}^{\dagger }b_{-k}^{\dagger }+b_{k}b_{-k})+v_{k}^{2}%
\big].  \label{eq:Ising_H_I}
\end{equation}%
Since the ground state is the vacuum state of $b_{k}$, the only contribution
to the spectral function is the term $4\sum_{k>0}iu_{k}v_{k}b_{k}^{\dagger
}b_{-k}^{\dagger }$. We find that the spectral function becomes%
\begin{equation}
I(0,\omega +i\eta )=\sum_{k>0}\frac{4\sin ^{2}k}{\epsilon (k)^{2}}\delta
\lbrack 4\epsilon (k)-\omega -i\eta ].
\end{equation}%
In terms of the Poisson kernel representation of the $\delta -$function, we
have
\begin{equation}
I(0,\omega +i\eta )=\sum_{k>0}\frac{\eta }{\pi }\frac{4\sin ^{2}k}{\epsilon
(k)^{2}}\frac{1}{\left[ 4\epsilon (k)-\omega \right] ^{2}+\eta ^{2}}.
\end{equation}

\begin{figure}[tbp]
\includegraphics[width=10cm]{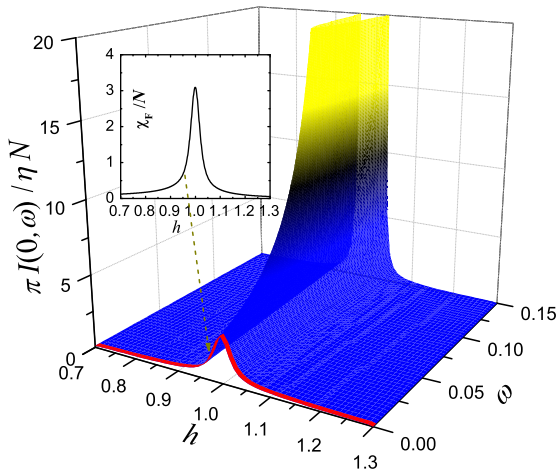}
\caption{(Color online) A 3D plot of the re-scaled spectral function as a
function of $\protect\omega$ and $h$. The inset is the fidelity
susceptibility as a function of $h$ for the Ising model, calculated from Eq.
(\protect\ref{eq:fsisingana}). Here $N=100$ and $\protect\eta=0.01$. The red
line on the 3D surface at $\protect\omega=0$ is consistent with fidelity
susceptibility in the inset, hence $\protect\chi _{F}\simeq \frac{\protect%
\pi }{\protect\eta }I(0,i\protect\eta )$ with $\protect\eta=0.01$.}
\label{figurfsspec.eps}
\end{figure}

In Fig. \ref{figurepec.eps}, we plotted the spectral function as a function
of energy $\omega $ and the driving parameter $h$ for a system of $N=100$
and $\eta =0.01$. According to the color scale definition, the bright region
denotes the resonance peaks of the spectral function. We can see that the
farther away from the critical point, the higher energy that the first peak
locates. This observation is consistent with the structure of the energy
spectrum of the quantum Ising model in which it is gapless only at the
critical point. To extract the fidelity susceptibility from the spectral
function, we plotted a 3D surface map of the spectral function $I(0,\omega
+i\eta )$ as a function of $\omega $ and $h$ for the same system and $\eta $
in Fig. \ref{figurfsspec.eps}. Clearly, though the spectral function becomes
smaller and smaller as the energy tends to zero, a line with sharp peak
appears in the cross-section of $\omega =0$. This line is the fidelity
susceptibility of the quantum Ising model. As a comparison, we reproduce the
fidelity susceptibility of the quantum Ising model in the inset of Fig. \ref%
{figurfsspec.eps}. These two lines are matched with each other. This fact
means%
\begin{equation}
\chi _{F}\simeq \left. \frac{\pi }{\eta }I(0,i\eta )\right\vert _{\eta =0.01}
\label{eq:fsspecdfdsfs}
\end{equation}%
for the present system. Therefore, by probing the spectral function of $%
A_{k} $, one can obtained the ground-state fidelity susceptibility which can
help us to find the critical point of the system. While Eqs. (\ref%
{eq:specinfidelitysus}) and (\ref{eq:specinfidelitysus2}) are more general
in physics, we need $I(0,i\eta )$ only to get the
fidelity susceptibility in quantum phase transitions. On the other hand,
since $A_{k}$ in Eq. (\ref{eq:specralfuudqs}) can be any driving operator of
many-body systems, as a more precise form of Eq. (\ref{eq:fsspecdfdsfs}),
the equality
\begin{equation}
\chi _{F}=\lim_{\eta \rightarrow 0}\frac{\pi }{\eta }I(0,i\eta )
\end{equation}%
is universally valid for any quantum phase transition.

In summary, we derived an interesting equality that relates the fidelity
susceptibility and spectral function in this work. Such an equality makes
it possible to measure the fidelity susceptibility directly in experiments
via the well known techniques, such as neutron scattering, ARPES
techniques, etc. Then we investigated the
feasibility of probing quantum criticality by measuring the fidelity
susceptibility in experiments. For this purpose, we take the one-dimensional
transverse-field Ising model as an example because the model can be realized
by the compound material CoNb$_{2}$O$_{6}$ in the laboratory . We show that
the fidelity susceptibility can be derived from the spectral function of the
driving operator of the model. Due to the important role of the fidelity in
detecting quantum phase transitions, we hope that equality will attract
experimentalists to study critical phenomena by measuring the fidelity
susceptibility directly in experiments.

SJGu thanks Ming Gong for helpful discussions. This work is supported by the
Earmarked Grant Research from the Research Grants Council of HKSAR, China,
under project CUHK 401212.


\begin{thebibliography}{99}
\bibitem{Sachdev} S. Sachdev, \textit{Quantum Phase Transitions}, (Cambridge
University Press, Cambridge, UK, 2000). %A book on QPTs

\bibitem{Nielsen1} M. A. Nilesen and I. L. Chuang, \textit{Quantum
Computation and Quantum Information} (Cambridge University Press, Cambridge,
England, 2000).

\bibitem{ZanardiPRE} P. Zanardi and N. Paunkovi{\' c}, Phys. Rev. E \textbf{%
74}, 031123 (2006).

\bibitem{HQZhouJPA2008} H. Q. Zhou and J. P. Barjaktarevic, J. Phys. A:
Math. Theor. \textbf{41}, 412001 (2008).

\bibitem{PZanardi0606130} P. Zanardi, M. Cozzini, and P. Giorda, J. Stat.
Mech. \textbf{2}, L02002 (2007).

\bibitem{MCozzini07} M. Cozzini, P. Giorda, and P. Zanardi, Phys. Rev. B,
\textbf{75}, 014439 (2007).

\bibitem{MCozzini072} M. Cozzini, R. Ionicioiu, and P. Zanardi, Phys. Rev.
B, \textbf{76}, 104420 (2007).

\bibitem{Buonsante1} P. Buonsante and A. Vezzani, Phys. Rev. Lett. \textbf{98%
}, 110601 (2007).

\bibitem{WLYou2007} W. L. You, Y. W. Li, and S. J. Gu, Phys. Rev. E \textbf{%
76}, 022101 (2007).

\bibitem{PZanardi2007G} P. Zanardi, P. Giorda and M. Cozzini, Phys. Rev.
Lett. \textbf{99}, 100603 (2007).

\bibitem{HQZhouPRL2008} H. Q. Zhou, R. Orus and G. Vidal, Phy. Rev. Lett.
\textbf{100}, 080601 (2008).

\bibitem{BogdanPRL2011} Marek M. Rams and Bogdan Damski, Phys. Rev. Lett.
\textbf{106} 055701 (2011).

\bibitem{SChenPRA2008} S. Chen, L. Wang, Y. Hao and Y. Wang, Phys. Rev. A
\textbf{77}, 032111 (2008).

\bibitem{VenutiPRL2007} L. C. Venuti and P. Zanardi, Phys. Rev. Lett.
\textbf{99}, 095701 (2007).

\bibitem{SJGuPRB2008} S. J. Gu, H. M. Kwok, W. Q. Ning and H. Q. Lin, Phys.
Rev. B \textbf{77}, 245109 (2008).

\bibitem{SJGUReview} S. J. Gu, Int. J. Mod. Phys. B \textbf{24}, 4371 (2010).

%topological phase transitions.

\bibitem{AHammaPRB2008} A. Hamma, W. Zhang, S. Haas and D. A. Lidar, Phys.
Rev. B \textbf{77}, 155111 (2008).

\bibitem{JHZhouPRB2009} J. H. Zhao and H. Q. Zhou, Phys. Rev. B \textbf{80},
014403 (2009).

\bibitem{SYangPRA2008} S. Yang, S. J. Gu, C. P. Sun and H. Q. Lin, Phys.
Rev. A \textbf{78}, 012304 (2008).

\bibitem{DFAbasto2008} D. F. Abasto, A. Hamma and P. Zanardi, Phys. Rev. A
\textbf{78}, 010301(R) (2008).

\bibitem{MFYangFidelity} T. C. Tzeng, H. H. Hung, Y. C. Chen and M. F. Yang,
Phys. Rev. A \textbf{77}, 062321 (2008).

\bibitem{HMKwok2008PRE} H. M. Kwok, W. Q. Ning, S. J. Gu and H. Q. Lin,
Phys. Rev. E \textbf{78}, 032103 (2008).

\bibitem{JieRenPRA} J. Ren, X. Xu, L. Gu, J. Li, Phys. Rev. A \textbf{86},
064301(2012).

\bibitem{GreschnerPRB2013} S. Greschner, A. K. Kolezhuk, and T. Vekua, Phys.
Rev. B \textbf{88}, 195101 (2013).

\bibitem{Oliveira2014} T. P. Oliveira and P. D. Sacramento, Phys. Rev. B
\textbf{89}, 094512 (2014); P. D. Sacramento, N. Paunkovi\'{o}, and V. R.
Vieira, Phys. Rev. A \textbf{84}, 062318 (2011).

\bibitem{WLYou2PRB} F. Trousselet, P. Horsch, A. Mle\'{e}, and W. L. You,
Phys. Rev. B \textbf{90}, 024404 (2014).

\bibitem{XuebingPRA2014} X. Luo, K. Zhou, W. Liu, Z. Liang, and Z. Zhang,
Phys. Rev. A \textbf{89}, 043612 (2014)

\bibitem{ZhangPRLexp} J. Zhang, X. Peng, N. Rajendran and D. Suter, Phys.
Rev. Lett. \textbf{100}, 100501 (2008).

\bibitem{ARPESbook} Riccardo Comin and Andrea Damascelli, \emph{ARPES: A
probe of electronic correlations} in the book \emph{Strongly Correlated
Systems: Experimental Techniques}, edited by A. Avella and F. Mancini,
Springer Series in Solid-State Sciences Vol. 180 (2013).

\bibitem{RColdeal} R. Coldea1, D. A. Tennant, E. M. Wheeler, E. Wawrzynska,
D. Prabhakaran, M. Telling, K. Habicht, P. Smeibidl, K. Kiefer, Science
\textbf{327}, 177 (2010).
\end{thebibliography}
\end{document}